\documentclass[10pt,letterpaper]{article}

\usepackage{amsmath,amssymb}

\usepackage{changepage}

\usepackage[utf8x]{inputenc}

\usepackage{textcomp,marvosym}

\usepackage{cite}

\usepackage{nameref,hyperref}

\usepackage{microtype}
\DisableLigatures[f]{encoding = *, family = * }

\usepackage[table]{xcolor}

\usepackage{array}

\makeatletter
\renewcommand{\@biblabel}[1]{\quad#1.}
\makeatother

\usepackage{lastpage,fancyhdr,graphicx}
\usepackage{epstopdf}
\begin{document}
\vspace*{0.2in}

% Title must be 250 characters or less.
{\Large
\textbf\newline{Ideological differences in engagement in public debate on Twitter} % Please use "sentence case" for title and headings (capitalize only the first word in a title (or heading), the first word in a subtitle (or subheading), and any proper nouns).
}
\newline
% Insert author names, affiliations and corresponding author email (do not include titles, positions, or degrees).
\\
Felix Gaisbauer\textsuperscript{1\Yinyang}*,
Armin Pournaki\textsuperscript{1\ddag},
Sven Banisch\textsuperscript{1\ddag},
Eckehard Olbrich\textsuperscript{1\ddag}
\\
\bigskip
\textbf{1} Max Planck Institute for Mathematics in the Sciences, Inselstra\ss e 22, 04103 Leipzig, Germany
\\
\bigskip

% Insert additional author notes using the symbols described below. Insert symbol callouts after author names as necessary.
% 
% Remove or comment out the author notes below if they aren't used.
%
% Primary Equal Contribution Note
\Yinyang These authors contributed equally to this work.

% Additional Equal Contribution Note
% Also use this double-dagger symbol for special authorship notes, such as senior authorship.
\ddag These authors also contributed equally to this work.

% Use the asterisk to denote corresponding authorship and provide email address in note below.
* Corresponding author, felix.gaisbauer@mis.mpg.de

% Please keep the abstract below 300 words
\section*{Abstract}
This article analyses public debate on Twitter via network representations of retweets and replies. We argue that tweets observable on Twitter have both a direct and mediated effect on the perception of public opinion. Through the interplay of the two networks, it is possible to identify potentially misleading representations of public opinion on the platform. The method is employed to observe public debate about two events: The Saxon state elections and violent riots in the city of Leipzig  in 2019. We show that in both cases, (i) different opinion groups exhibit different propensities to get involved in debate, and therefore have unequal impact on public opinion. Users retweeting far-right parties and politicians are significantly more active, hence their positions are disproportionately visible. (ii) Said users act significantly more confrontational in the sense that they reply mostly to users from different groups, while the contrary is not the case.

% Use "Eq" instead of "Equation" for equation citations.
\section*{Introduction}
Twitter is an immensely popular object of study for social scientists in a variety of contexts, ranging from politics  \cite{jungherr2015} to crisis communication \cite{bruns2014crisis}. One reason for this popularity is that Twitter is open in a double sense: On the one hand, researchers can call Twitter data conveniently via an API. On the other hand---and more importantly---the content created on the platform is public by default. \textit{In principle}, a user's activity is visible to everyone on the platform, and any user can interact with anyone else.

Due to its open platform design, user interactions on Twitter might, of all major social media platforms, come closest to what is commonly referred to as `public debate.' While not being representative of the general public \cite{mellon2017twitter,mislove2011}, Twitter provides a public arena for information gathering, opinion formation and persuasion. Since journalists incorporate the platform in their daily routines \cite{broersma2013,paulussen2014}, explicitly refer to content visible on Twitter as public opinion \cite{mcgregor2019social}, and even tend to judge tweets as newsworthy as press agency reports \cite{mcgregor2020twitter}, the standpoints that are prominently featured there are reinforced in traditional media. A better understanding of how different opinion groups shape debate on the platform is therefore highly important: The image created on the platform not only affects how public opinion on certain issues is perceived by its users, but by society more generally. Certain standpoints---advanced by committed minorities in particular---might appear more prevalent than they actually are. 

This explorative study attempts to make these systematic differences in the engagement of groups with different political leaning visible. Its goal is both methodological and case-oriented: First, we propose a novel method to assess what users and lurkers perceive as public opinion on a specific issue on Twitter. Secondly, we employ this method in two case studies covering two political events in Germany.

To this end, we firstly choose a suitable theoretic underpinning for the concepts of public debate and public opinion, which have been interpreted from different angles \cite{noelle1984,price1992}, and which we connect to findings on user comments and their effects on readers. 
Then, we describe how relevant tweets are collected, and the data transformations which yield a social-structural view on interactions between users. The proposed method relies on an interplay of network representations of two types of user interactions on Twitter: Retweets and replies. Retweet networks are used to discern opinion groups, while reply networks make it possible to assess how these groups participate in public debate.
We construct said networks from a user-centered data collection for two events: A state election in the German state of Saxony and a violent clash on New Year's Eve between police and parts of the population in the city of Leipzig. We show for both cases that while retweet networks are strongly polarized, debate between users of different opinion clusters is vivid. We also show that the impression of public opinion is biased. While being a minority in number, Twitter users who mostly share content of parties to the right of the political spectrum are disproportionately active in debate and act more confrontational in the sense that they address users from different opinion groups more often.

\section*{Theory}
\subsection*{Public debate and public opinion}
A communication-based view on public debate has been provided by Vincent Price. He describes public debate as ``communication processes through which publics are constituted and within which opinions on public affairs are formed'' \cite[~p. 74]{price1992}. While he invokes the analogy of a big town meeting, the technical feasibility of creating such a meeting still seemed out of reach in the early nineties: ``Modern communication technologies may have enabled the enlargement of public consciousness [...] but they have not come close to creating any sort of town meeting at large''  \cite[~p. 78]{price1992}. 

With the advent of social media, the analogy appears to have become a (digital) reality. As has been stressed in the introduction, Twitter is a public medium and allows its users to interact with potentially anyone else on the platform. Users can share others' thoughts, put out their own, and organize around hashtags, thereby creating publics and attracting attention of others \cite{bruns2015twitter}. These processes are strongly reinforced and amplified by traditional media: Journalists incorporate social media, especially Twitter, as an established news source in their daily routine \cite{paulussen2014,broersma2013,mcgregor2020twitter}, journalistic content or events on television are discussed on the platform in parallel \cite{trilling2015,gearhart2014}, and Twitter content is often explicitly used to represent public opinion, both in qualitative (by quoting certain tweets, e.g. to underline meta-narratives) and quantitative fashion \cite{mcgregor2019social}. Often, social media platforms themselves provide tools or even supply journalists with data or analyses in order to get mentioned in their articles \cite{mcgregor2019social}.

Two different, basic paradigms of public opinion can be subsumed under the terms \textit{discursive} and \textit{demoscopic} public opinion \cite{price1992,schweiger2008}. The former refers to public opinion as a social-structural phenomenon and has a strong normative imprint. The process of arriving at public opinion---public debate---is understood as as a rational discourse between well-informed citizens \cite{schweiger2008}, and should lead to the best possible decision with respect to the overall good. The latter is related to survey research where scientists seek to aggregate the attitudes of individuals towards certain issues in a representative fashion, which yields, by majority rule or a breakdown of percentages, public opinion.

For an understanding of online interactions, both conceptions are problematic. There have been attempts to replace classical voter surveys for elections by social media observations \cite{tumasjan2010,burnap2016ukelection}, but the findings have been contested by others \cite{jungherr2015} or turned out to be incorrect (Burnap et al. \cite{burnap2016ukelection} predicted a Labour win in the 2015 UK elections). Discursive public opinion as a normative concept, on the other hand, is in general hardly accessible to empirical research. And, after all, the internet does know many compulsions besides ``the unforced force of the better argument'' \cite[~p. 306]{habermas2015}. 

Elisabeth Noelle-Neumann has proposed a social-psychological approach centered around \textit{observable} opinion expressions. (`Expressions' is interpreted very broadly: Noelle-Neumann's account includes also non-verbal modes of communication, e.g. badges that support certain political parties or even subtle facial expressions such as raised eyebrows.) She conceives public opinion---or rather, what people perceive as the opinion of `the' public---as a force of informal social pressure and control manifesting itself in ``approval and disapproval of publicly observable positions and behavior'' \cite[~p. 64]{noelle1984}. Her operational definition of public opinion incorporates ``opinions on controversial issues that one can express in public without isolating oneself'' \cite[~p. 63]{noelle1984}. Especially for controversial topics, individuals, being social creatures and fearing social isolation, constantly and mostly sub-consciously monitor their social environment and the mass media. They estimate the majority opinion around them, employing some ``quasi-statistical sense'' \cite{noelle2004}, which they then refer to as public or popular opinion. The theory hence puts strong emphasis on the role of the subjective impression of public opinion of individuals. 
Noelle-Neumann's spiral of silence theory states that if people realize that they hold an opinion that differs from their impression of public opinion, they tend to be less willing to express their opinion publicly any longer. This, in turn, affects the perception of public opinion of others, potentially setting off a spiralling process in which certain groups become more expressive while others fall silent.
Quantified interactions in online environments (Twitter displays the number of likes, retweets, and replies below each tweet) might suggest themselves as an objective foundation for the quasi-statistical impression of the opinion climate---but as we will show, these interactions themselves can be subject to strong biases.

\subsection*{Comment spaces and perceived public opinion}
In order to capture public debate online, comment sections, predominantly on news websites, have been the target of attention since their introduction \cite{springer2018,lee2010,lee2012,friemel2015,toepfl2015}. Studies found that user comments on news articles affected individuals' perceptions of public opinion \cite{lee2010,lee2012}---more so than simple comparisons of likes and dislikes. 
In the 20th century, scholars often distinguished between representatives of interest groups that debated publicly, and a large and more spectator-like `body' which then reacted to the debate and approved or disapproved, hence formed public opinion \cite[~p. 27]{price1992}. This relation is reproduced by the combination of newspaper articles and user comments below, which allow directly visible engagement in larger amount and with less control than the very limited number of redacted letters to the editor.

Allowing differing points of view to reach an audience with few formal constraints, comment spaces have also been interpreted as counter-public spaces (or spheres) \cite{toepfl2015,kaiser2017}, an expansion of the Habermasian concept of the public sphere. Nancy Fraser originally defined counterpublics as ``parallel discursive arenas where members of subordinated social groups invent and circulate counterdiscourses to formulate oppositional interpretations of their identities, interests, and needs'' \cite[~p. 67]{fraser1990}, arising in response to hegemonic ``publics at large'' \cite[~p. 67]{fraser1990}. Comment sections in general (may they be the comment sections of newspapers, or the reply thread of a tweet of a public figure on Twitter) are quite suitable for the formulation of oppositional interpretations: They are in the direct vicinity, nevertheless clearly demarcated from the interpretations and content they want to distance themselves from \cite{toepfl2015}. 

Lessons of these findings and interpretations are (i) that comment sections have a significant effect on how people perceive public opinion on an issue---in Noelle-Neumann's terms, the ``opinion climate'' \cite{noelle1984}---and (ii) that they are hence, for all kinds of interest groups, important arenas for confrontation and contestation of certain interpretations and frames, may they be hegemonic or not. Therefore, a careful investigation of comment sections and the views expressed there is important: Which standpoints are expressed, how often, and which viewpoints (or users) remain silent? That different groups strive for the award of being called \textit{the} public is nothing new. As Baker notes of the pre-revolutionary times in France: 
``Indeed, one can understand the conflicts of the Pre-Revolution as a series of struggles to fix the sociological referent of the concept in favor of one or another competing group'' \cite[~p. 186]{baker_1990}. Online environments, which facilitate communication and decentralize information distribution, might appear to make this competition more transparent. But they also introduce additional potential for misperceptions, not least due to differing willingness of public opinion expression of different groups \cite{kalogeropoulos2017,friemel2015,mustafaraj2011}. For certain opinion groups, this can lead to a False Consensus Effect \cite{ross1977false}, according to which individuals see their own opinions as more prevalent in society than they actually are. On the other hand, groups less willing to express their opinion might underestimate their size (False Uniqueness Effect, see Mullen, Dovidio, Johnson and Copper \cite{mullen1992group}). The method in this contribution will allow an estimation of how public opinion and public debate are perceived on Twitter, and which opinion groups principally shape this impression.

\section*{Political background}
The two events under consideration were the Saxon state election which took place on September 1st, 2019, and a violent clash between police and parts of the population in the city of Leipzig on New Year's Eve four months later (in the following abbreviated with NYE). The events were complementary in the sense that the election was long-anticipated, while the latter occasion was a spontaneous incident, making them suitable for comparison.

The election was of special, nation-wide interest. Since Saxony had been the birthplace of the anti-Islam movement \textit{Pegida} in 2014, which received international attention, the election was considered a litmus test for the mobilizing potential of extreme forces. 
Before the election, it was not clear whether there would be the possibility of forming a majority coalition without participation of the \textit{Alternative für Deutschland} (\textit{AfD}), a right-wing party founded in 2013, or the democratic socialist party \textit{Die Linke}. While the Christian democratic party \textit{Christlich Demokratische Union Deutschlands} (\textit{CDU}) leading the polls ruled out coalitions with both of the parties, it was publicly discussed whether parts of the \textit{CDU} were open towards collaboration with the \textit{AfD} \cite{mdr}.

The NYE incident was not anticipated, but a spontaneous event which was subsequently discussed not only in Saxony, but also in national politics. On New Year's Eve, violent riots and attacks on police officers occurred in the city of Leipzig's quarter Connewitz. The event was particularly polarizing:
While some political actors framed the incident as an example of the violent potential of left-wing extremism, others accused the police of deliberate provocation \cite{tagesspiegel}.

\section*{Methods}
\subsection*{Data acquisition}
The data was collected in a user-centered approach. That is, all tweets that were produced by a seed set of users were gathered. Moreover, all tweets containing the Twitter handle of one or more of the seed users were collected---this included retweets, mentions, and replies to the users. With this method, not only first-order replies to a user in question could be collected, but any part of a reply tree that had been initiated by the user. 
In the months preceding the Saxon state elections in September 2019, a first seed set was constructed. Candidates in elections in Germany appear on electoral lists. We collected all names from these lists and checked whether the candidates had an actively maintained Twitter account. If this was the case, we included them. We also included \begin{itemize}
    \item state party accounts
    \item the leaders of the fractions in the state parliament and the local party organizations
    \item the Saxon members of the national parliament
    \item the Saxon members of the European parliament
    \item Saxon media accounts
    \item Saxony correspondents of national media
\end{itemize}
that had an active Twitter account.
The seed set was expanded with a snowball-sampling method: In each sampling step, users that were not contained in the seed set but retweeted or mentioned at least once a week by users that were already in the seed set and were related to Saxony were included. The latter criterion was necessary to exclude nation-wide media accounts and national politicians.  At the end of July, after seven iterations, the final seed set consisted of 270 users.

Tweets were gathered until February 2020. This allowed observation of other Saxony-related events, such as the NYE incident. Since the seed set was not perfectly tailored towards this event, we restricted the analysis to the subset of tweets containing certain event-specific keywords (\textit{connewitz, antifa, polizei, polizist, le0101, linx, leipzig, Not-OP, notop, linke, chaoten, angriff, le3112, randal,} applied to the root tweets and incident-specific retweet network).
The tweets used for the analysis of the election stem from the time period between the 25th of July and the 10th of September (364,626 tweets). For the NYE data set, tweets from December 31 until January 19 (130,685) were used. 
The two cases were chosen since one (NYE) represents a spontaneous event, while the other a long-anticipated election---a suitable test whether similar effects can be observed for both types of events. Moreover, the data sets were of considerable, but still maintainable size (especially regarding visualization) and therefore appropriate for testing the method.
\subsection*{Network representations}
We used networks as a mathematical abstraction to represent two types of interaction in the data set: retweets and replies. Retweet networks were used to discern different opinion communities on Twitter, while reply networks made it possible to assess how these groups participated in public debate.

\subsubsection*{Retweet networks}
Retweet interactions are represented as a directed network in which every node is a user. A link is drawn from user $a$ to user $b$ every time $a$ retweets $b$.
It has become standard practice to employ community detection algorithms to find strongly connected clusters in a retweet  network which are then interpreted as groups of users sharing an opinion or political position \cite{Conover2011,conover2011polarization,Gaumont2018}. 

An alternative approach is the spatialisation of the retweet network via a force-directed algorithm, such as ForceAtlas2 \cite{jacomy2014}, which we used in its Gephi \cite{bastian2009gephi} implementation. Noack \cite{noack2009} has shown that the energy-minimal states of force-directed layout algorithms such as ForceAtlas2 are in fact relaxations of modularity maximization. While in order to maximize modularity, nodes must be classified into discrete clusters, force-directed layouts assign continuous positions---in the case of ForceAtlas2, in a two-dimensional space.
Positions in the layout are hence closely related to partition outputs of modularity maximization algorithms such as Louvain. The advantage of force-directed layout algorithms is that one can \textit{visually} distinguish tightly-knit clusters and less dense in-between regions, which might be classified as one cluster in the Louvain algorithm. While ``community detection algorithms tend to generate clear-cut and non-overlapping partitions, force-directed spatialisation reveals zones of different relational density but with blurred and uncertain borders'' \cite[~p. 13]{venturini2019}. Public debate and the different opinion camps cannot be clearly demarcated from one another, and regions of transition between different groups, that do not clearly belong to any one of them, are politically meaningful and should hence be discernible. As will be shown in the following section, retweet networks of both events showed polarized structure in the force-directed layout. We classified the two cohesive poles of the retweet network as two opinion clusters (the borders chosen as is visible in Fig \ref{fig:rtnws} A and B), and assigned the in-between region to a different cluster upon visual inspection. 
%This will make the analysis more coarse-grained in one sense, but more nuanced in another: There will usually be fewer communities, but the communities will represent more fundamental divides and it will be possible to discern the border regions in-between, which can be treated as separate clusters.

% The users were classified into one of three opinion camps with respect to their position in this polarized layout (one for the pole which included the majority of users, one for the minority pole, and one for the users in the border region between the poles).

\subsubsection*{Reply trees and reply networks}

Due to the user-centered data collection, it was also possible to retrieve an exhaustive collection of all replies that were initiated by posts of the seed users. 
A post together with all its replies can be represented by a reply tree (see Fig~\ref{fig:tree_example}). Only taking into account reply trees initiated by the prominent seed users corresponds quite naturally to the distinction of the previous section between representatives of interest groups debating publicly and the spectator-like body approving or disapproving subsequently \cite{price1992}. These reply threads then function as spaces where different opinion groups can confront each other: They are widely visible due to the prominence of the creator of the tweet which spans up discussion, and can hence attract users of different opinion camps. Retweets, on the other hand, mainly serve to share information with one's followers. A retweet might point towards a debate, but does not imply involvement in it.

In order to gain a global view on public debate, we aggregated the combined interaction structure of all reply trees into one reply network, assigning a directed edge between two users if one had directly replied to the other in a tree (see Fig~\ref{fig:rtnws}). (Obviously, trees are networks, too. But if we in the following speak of reply networks, we mean the bigger networks constructed in this procedure.) 

Some works have taken similar routes by taking into account direct user interactions in the form of mentions \cite{conover2011polarization} and replies \cite{sousa2010,aragon2013,yardi2010}. Sousa, Sarmento and Mendes Rodrigues \cite{sousa2010} and Yardi and boyd \cite{yardi2010} use a keyword-based tweet collection. This approach is useful if one is solely interested in tweets that include a certain keyword, while full conversations in a reply thread between users are not accessible with the method. Aragón, Kappler, Kaltenbrunner, Laniado and Volkovich \cite{aragon2013} and Nuernbergk and Conrad \cite{nuernbergk2016} employ a user-centered collection and construct a reply network, but only between politicians on Twitter and hence do not capture debate among a more general public. Since the data sets in the present contribution include the complete reply trees below each post of one of the seed users, it was possible to gain a more general perspective on public debate that did not only include certain elites. 

The classification of users from the retweet network---the information about whether they belonged to the minority or majority pole, or the in-between region---was imported into the reply trees and networks.
This made it possible to investigate how many users of the different retweet clusters were also involved in public debate, hence willing to express their opinion in discussion with others of possibly different opinions, and whether users of different opinion clusters debated mainly among each other or with others. 

It must be noted here that not all users involved in debate were present in the retweet network.
Hence, the classification in the reply trees and networks was not complete. Initially, around 47\% of the users involved in debate in the election data set could be classified (33\% for NYE). In order to include more users in the classifications, a larger retweet network was additionally constructed which included all retweets from July 2019 until the end of February 2020. The overall structure of the network was similar to the incident-specific retweet networks (see Supplementary Material). If a user was present in the reply trees, but not present in the incident-specific retweet network, it was checked whether the user was present in the large retweet network---if so, the user was assigned the classification from this network. With the use of the big retweet network, 63\% (election) and 67\% (NYE) of users present in the reply trees could be classified.

\section*{Results}
\subsection*{Retweet networks and classification}
\begin{figure*}[!h]
\begin{adjustwidth}{-1in}{0in}
\includegraphics[width = 16cm]{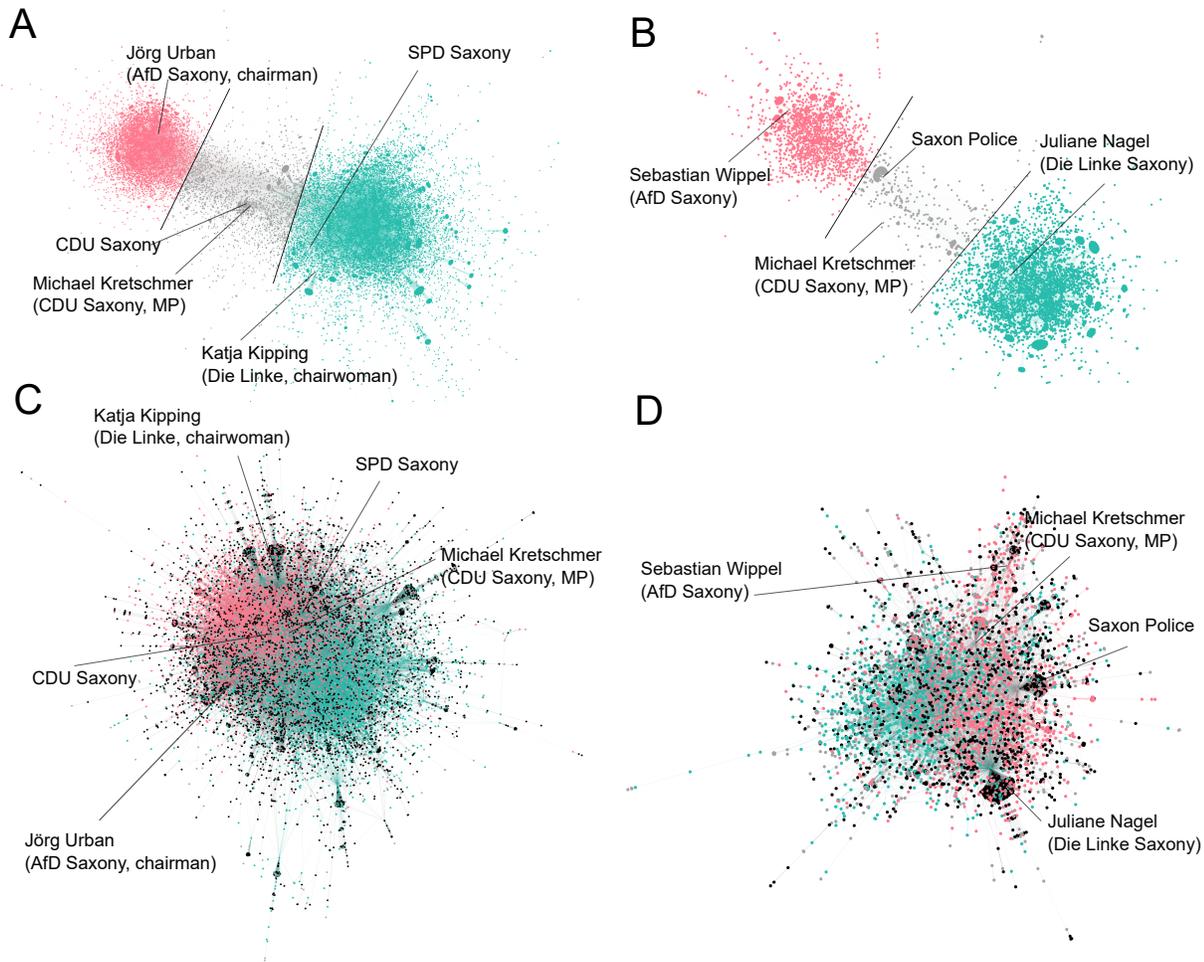}

\end{adjustwidth}
\caption{\textbf{Retweet and reply networks.} The giant component of the retweet networks (above) for the elections (A) and the NYE incident (B), and the giant component of the reply networks (C: elections, D: NYE). The retweet networks are both polarized, and users are classified according to their position in the force-directed layout. The borders for the classifications are included (black lines). The classifications are imported in the reply networks, which are not polarized. Black nodes indicate users that do not show up in the corresponding retweet network.\label{fig:rtnws}}

\end{figure*}

The retweet networks for both cases are strongly polarized (see Fig~\ref{fig:rtnws} A/B) in the force-directed layout. 

For the election data set (31,108 users in the giant component), seed users placed in the majority pole are politicians of the parties \textit{SPD}, \textit{Die Linke}, and \textit{Bündnis 90/Die Grünen} (and one politician of the \textit{CDU}), along with media accounts (e.g. \textit{Bild Leipzig}, \textit{LVZ}, or \textit{MDR Sachsen}) and left-wing activists. In the region between the two clusters, politicians of the \textit{CDU}, \textit{Freie Waehler} and \textit{FDP} are placed, as well as media accounts (e.g. \textit{MDR Aktuell}, \textit{Bild Dresden}, or \textit{TAG24}). The minority pole, on the other hand, includes seed users from the \textit{AfD}, \textit{Freie Waehler}, \textit{Blaue Partei} and the anti-Islam movement \textit{Pegida}.
The structure of the retweet network hence quite accurately mirrors the political constellations in the run-up to the elections. Left-wing and eco-friendly parties are placed in one cluster and the right to far-right parties in another, while politicians of the market-liberal \textit{FDP} and the center-right \textit{CDU} are located in-between the two.

The users of the majority cluster made up 64.5\% (20,052) of the retweet network, 23.1\% (7,195) were part of the minority cluster, and 12.4\% (3,861) of the users were in the region between the two.

A very similar structure, both in proportions and in political leanings, is given for the NYE incident. Some differences occur, however. The set of users placed in-between the two clusters (711 or 7.9\%) includes the official account of the city of Leipzig and the account of the Saxon police, as well as one politician of the \textit{AfD}, one from \textit{Die Linke} and one \textit{SPD} politician. The majority (6,010 users, 66.6\%) and minority cluster (2,301 users, 25.5\% of the giant component) show similar composition as in the election retweet network.

The classification of users on the basis of their position in the force-directed layout of the retweet network was taken as a proxy for the political position of the users in the two issues. It must be noted, however, that users of one cluster should not necessarily be interpreted as holding exactly the same opinion or political position. Rather, the clusters reflect an issue-specific fundamental political difference which is then also reflected by the classification.

A randomly selected subset of 100 users was used check whether the groups assigned to the users were plausible in the sense that the users tweeted content sympathetic to one of the parties or political figures in their assigned cluster. Out of the 100 users, 96 acted consistently with their classification, while 4 did not.

\subsection*{Reply trees and engagement}
\begin{figure*}
\begin{adjustwidth}{-1in}{0in}
    \includegraphics[width = 16cm]{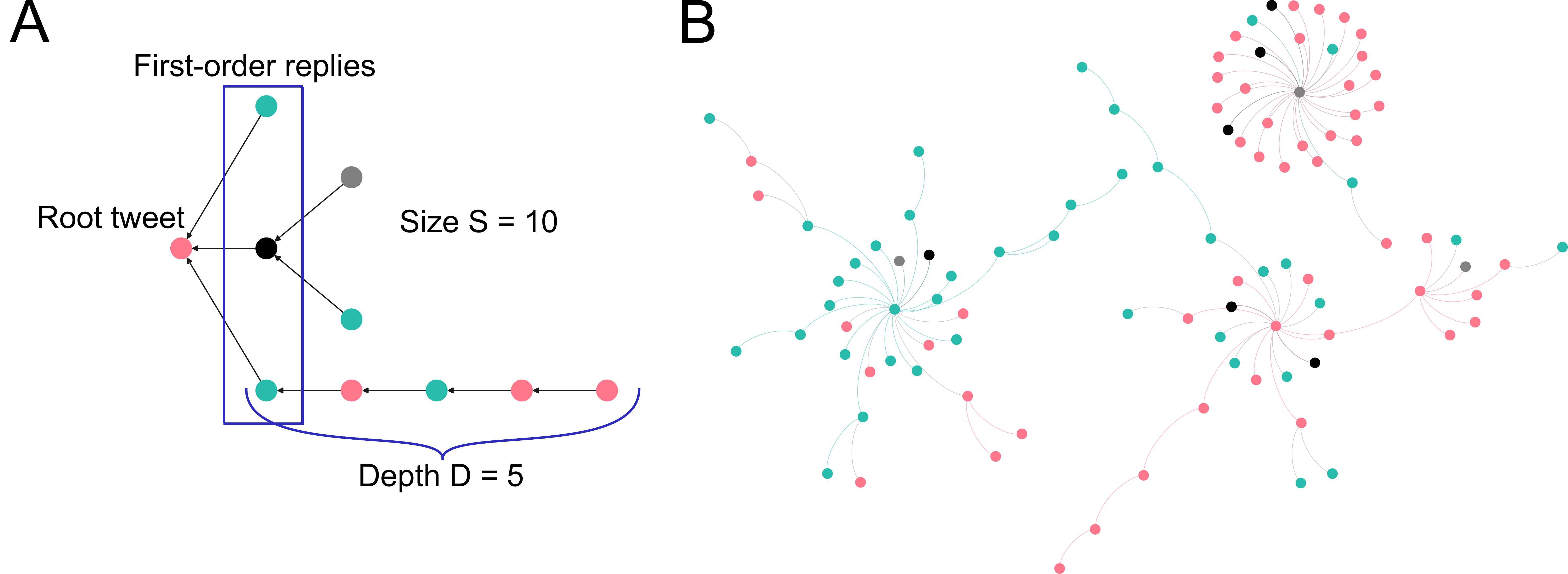}
\end{adjustwidth}
    \caption{\textbf{Reply trees.} An exemplary dummy reply tree (A), along with two reply trees from the data set (B). Each node represents a tweet, and a directed edge between two nodes indicates a reply. If the users of the tweets appear in the retweet network, their replies were color-coded according to the cluster of the user (a black node indicates a reply by a user that does not appear in the retweet network).
    The root tweet is the original tweet by one of the seed users, while first-order replies are the direct replies to the root tweet.}
    \label{fig:tree_example}

\end{figure*}

All reply trees that had been initiated by a root tweet of one of the seed users were taken into account in the analysis of the reply interactions. The reply trees themselves can be seen as representations of discussions triggered by single statements, and can exhibit arbitrarily complicated tree shape (see Figure \ref{fig:tree_example}) and have arbitrarily many participants. Not every tweet receives replies. There are 23,221 posts from the seed users in the election data set that were not replies or retweets, out of which 8,033 received at least one reply. (NYE: 2,020 posts, 897 with at least one reply.)

\begin{figure*}
\begin{adjustwidth}{-1in}{0in}
\centering
    \includegraphics[width = 16cm]{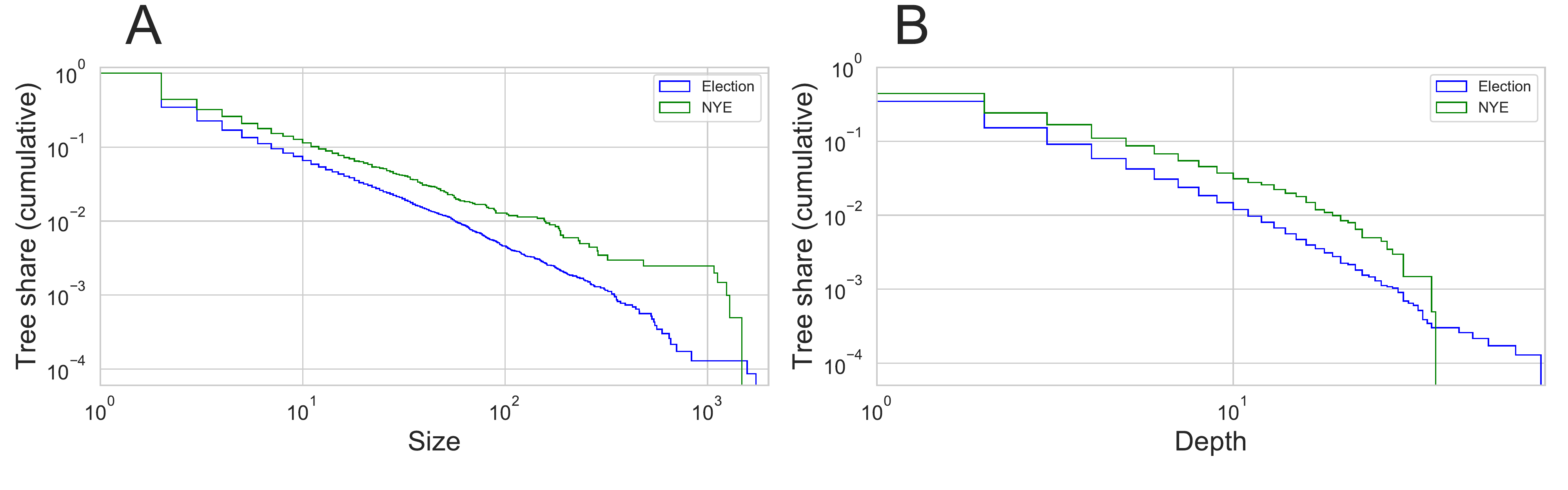}
\end{adjustwidth}

    \caption{\textbf{Reply tree size and depth.} Reply tree size (A) and depth distribution (B) for the two events.}
    \label{fig:tree_histograms}

\end{figure*}

Reply trees can be characterized by two quantities: Their size $S$, which is the overall number of tweets in the tree, and their depth $D$, which is the longest branch of the tree (Fig~\ref{fig:tree_example}). Fig~\ref{fig:tree_histograms} shows the cumulative distribution of sizes and depths of the reply trees in the two data sets. In both data sets, around 90 percent of all reply trees have a size smaller than 10 and a depth smaller than 5. Nevertheless, reply trees can be very large: The largest tree in the election data set has 1,936 replies (NYE: 1,475), and the maximum depths are 72 and 37, respectively.


By importing the cluster classification from the respective retweet network, it was possible to compare the engagement of the different groups in the reply trees. The engagement proportions in the debate differ strongly from those in the retweet network, both in number of users and number of tweets (see Table \ref{tab:repl+users}).

\begin{table*}[!ht]
\small
\centering
    \caption{Users and replies from the different retweet clusters involved in the reply trees.}    \label{tab:repl+users}
\vspace{0.3cm}
\begin{tabular}{lllll}
\hline
              & \multicolumn{2}{c}{Election} & \multicolumn{2}{c}{NYE} \\ 
              & Users        & Replies       & Users     & Replies     \\ 
              \hline
Majority cl.   & 6,736 (33.3\%)        & 30,615 (40.5\%)        & 2,008 (32.4\%)     & 6,403 (39.2\%)       \\
Minority cl.   & 4,696 (23.2\%)        & 23,790 (31.5\%)       & 1,727 (27.9\%)     & 4,858 (29.8\%)      \\
Intermediate cl. & 1,470 (7.0\%)        & 6,785 (9.0\%)       & 389 (6.3\%)    & 1,235 (7.6\%)      \\
Not classified     & 7,395 (36.5\%)       & 14,344 (19.0\%)       & 2,072 (33.4\%)     & 3,816 (23.4\%)     \\
\hline
\end{tabular}
\end{table*}

In the election data set, participants from the majority and minority retweet clusters made up 33.3\% and 23.2\% of all replies, respectively. For the NYE incident, even more users from the minority retweet cluster participated in the debate---32.4\% belonged the majority in the NYE retweet cluster, while 27.9\% were minority users involved in the debate (see Table \ref{tab:repl+users}). Users placed in-between the two poles in the retweet network also participated in the debate, but less so both in number of users and in number of replies.
36.5\% (election) and 33.4\% (NYE) of the users that participated in the reply trees were not present in the retweet network (i.e., they did not retweet any of the seed users nor any tweets that contained the Twitter handle of one of the seed users). Interestingly, while these users make up the biggest number of users involved in the debate, they do not constitute the majority in terms of replies. In both data sets, users from the poles of the retweet network, if involved in the debate, are most active. Users from `outside' do not tend to debate often and extensively---on average, in both data sets, they only give around two replies. 

In comparison, the ratio between majority and minority pole in the corresponding retweet network was 64.5\% to 23.1\% (election) and 66.6\% to 25.5\% (NYE). 
\begin{table}[!ht]

\centering
    \caption{First-order replies by retweet clusters. First-order replies from users of the two poles are roughly equal in number in both data sets and make up the majority of all first-order replies. The minority cluster is even more active in replies of first order than in reply trees in general. }    \label{tab:firstorder}
\vspace{0.3cm}
\begin{tabular}{lll}
\hline
 & Election & NYE \\ 
 \hline
Majority cl. & 15,347 (36.0\%) & 2,445 (32.7\% )\\
Minority cl. & 14,911 (35.0\%) & 2,451 (32.8\%) \\
Intermediate cl. & 3,710 (8.7\%) & 572 (7.6\%) \\
Not classified & 8,629 (20.3\%) & 2,014 (26.9\%)\\
 \hline
\end{tabular}
\end{table}

First-order replies are of special interest since they are usually directly displayed below the root tweet on Twitter. Therefore, they most probably have a stronger impact on the perception of public opinion than tweets that are at the end of a long discussion branch. The amount of first-order replies by the different clusters is displayed in Table \ref{tab:firstorder}. First-order replies from users of the two poles are roughly equal in number in both data sets. Minority pole users hence produced an even larger proportion of highly visible replies. Users from the intermediate region in the retweet network only account for less than 10 percent of first-order replies, while users that were not present in the retweet networks produced 20.3\% (election) and 26.9\% (NYE) of the replies of first order.

\begin{table}[]
\small
\centering
 \caption{Percentage of users in the incident-specific retweet networks that are active in the reply networks (seed users excluded) by cluster. The share of users from the minority pole is, in both cases, around twice as big as the share of users from the majority pole. User share from the in-between region is slightly bigger than that of the majority pole in both cases.}  \label{tab:active perc}
 \vspace{0.3cm}
\begin{tabular}{lll}
\hline
& Election & NYE  \\ 
\hline
Majority cl. & 22.4\%  (4,644)&  16.2\% (977) \\
Minority cl.  &  48.2\%  (3,604)&32.2\%  (752) \\
Intermediate cl. &   26.4\% (1,050) &   22.0\% (707) \\
\hline
\end{tabular}
\end{table}

Comparing engagement in the form of replies and retweets makes it possible to assess whether the different opinion groups show different inclination to participate in the debate. To this end, we calculate the share of users present in the different clusters of the incident-specific retweet networks  that were also present in the respective reply network (see Table \ref{tab:active perc}). For both events, the groups showed significantly different behavior (election: $\chi^2 = 850.7$, $p<0.001$; NYE: $\chi^2 = 138.2$, $p<0.001$). Users of the minority cluster were roughly twice as likely to get involved in the debate than users belonging to the majority cluster (election: z-score 41.0, $p<0.001$; NYE: z-score 15.0, $p<0.001$). Users from the in-between region were slightly more active in the debate than users from the majority pole (election: z-score 5.4, $p<0.001$; NYE: z-score 4.0, $p<0.001$), but still less than the minority pole (election: z-score $22.1$, $p<0.001$; NYE: z-score 46.0, $p<0.001$).

Hence, two findings are worth stressing: (i) Minority pole users are disproportionately active in the debate compared to the majority pole, both in number of users involved and of replies written (see also the Discussion). This effect is even more pronounced in first-order replies that are, by platform design, most visible. And (ii), users from both poles of the retweet network, if they take part in the debate in the form of replies, do so more extensively than users from in-between the poles or unclassified users. 

\subsection*{Reply networks and global interaction patterns}
The reply networks give a more comprehensive structural picture of debate---it is possible to make visible patterns of discussion between different users and user groups beyond interactions in single reply trees. Each reply network was constructed by aggregating all reply interactions in the reply trees into one big network, where each node represents a user and a directed edge is created between two users if one has replied to the other. 

The question of interest here is whether the groups also exhibit large-scale polarization when they discuss among each other, and whether there are differences in discussion behavior between the groups.
If public debate was fragmented in the sense that discussion ties were only existent amongst a certain subset of users, this would be visible in the force-directed layout of the reply network. But, as is displayed in Fig~\ref{fig:rtnws} C/D, this is the case for neither of the two events (again, spatialisation was carried out with ForceAtlas2). Users of clusters that are clearly separated in the force-directed layout of the retweet networks interact quite frequently in the form of replies.

\begin{figure*}
\begin{adjustwidth}{-1in}{0in}
\centering
    \includegraphics[width = 16cm]{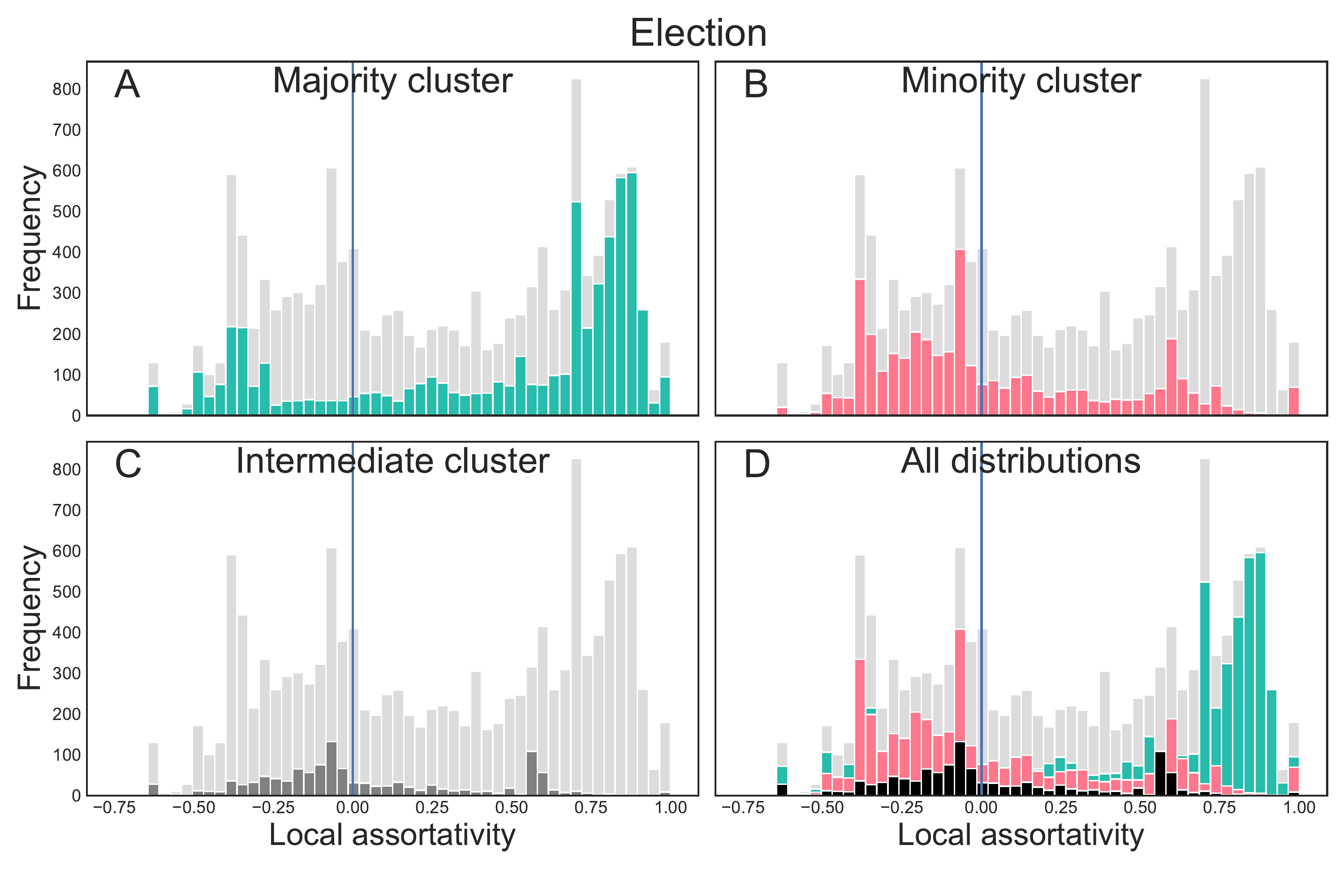}
\end{adjustwidth}

   \caption{\textbf{Local assortativity distribution for the election data.} Local assortativity distribution for the reply networks of the election, split up by the users' respective retweet clusters (A: majority cluster, B: minority cluster, C: intermediate cluster, D: comparison of all distributions). The grey distribution in the background is the overall local assortativity distribution, the distribution of the respective subgroup(s) is plotted in the foreground and color-coded as before.}
    \label{fig:local_ass_el}
\end{figure*}

\begin{figure*}
\begin{adjustwidth}{-1in}{0in}
\centering
    \includegraphics[width = 16cm]{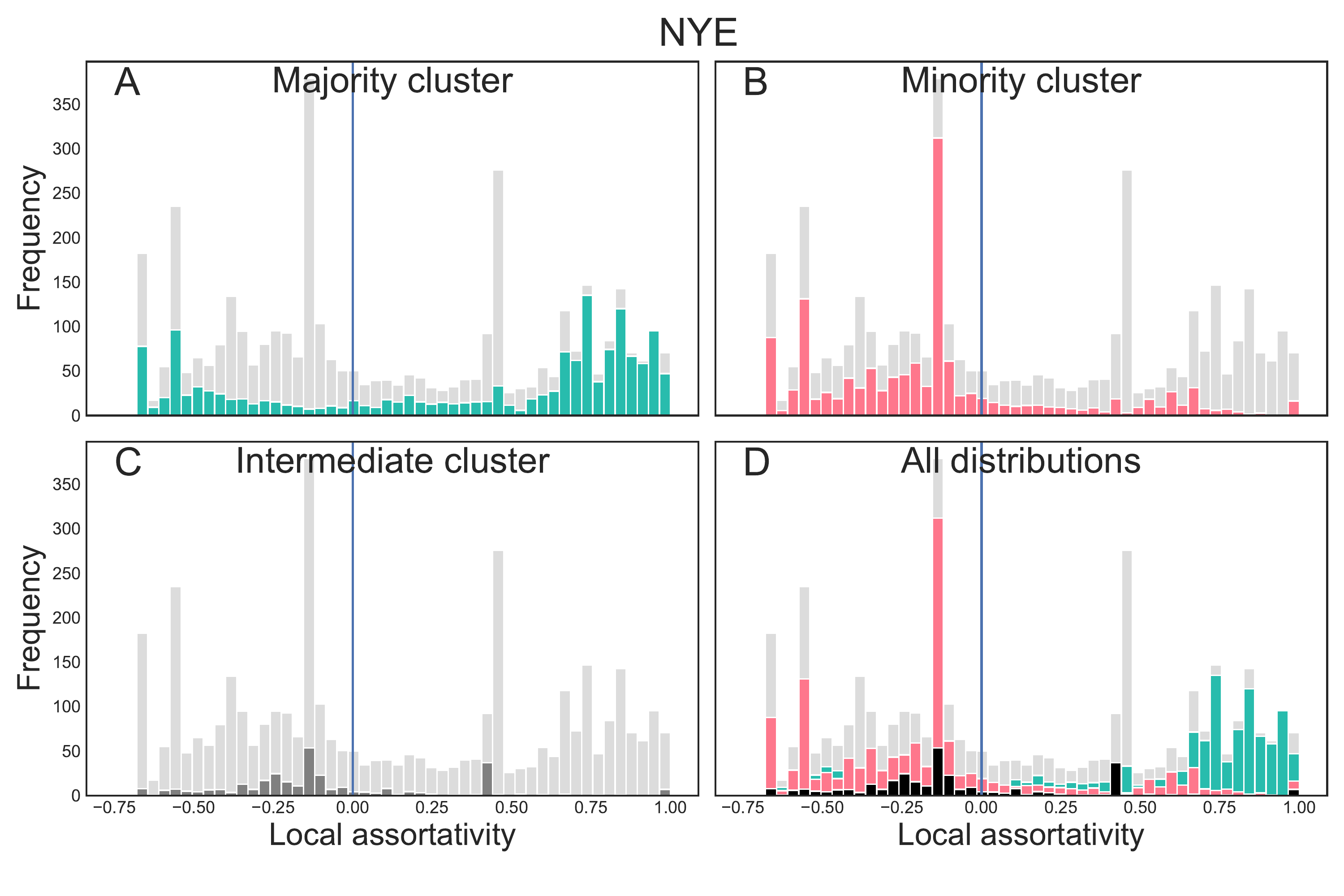}
\end{adjustwidth}

   \caption{\textbf{Local assortativity distribution for the NYE data.} Local assortativity distribution for the reply networks of the NYE incident, split up by the users' respective retweet clusters (A: majority cluster, B: minority cluster, C: intermediate cluster, D: comparison of all distributions). The grey distribution in the background is the overall local assortativity distribution, the distribution of the respective subgroup(s) is plotted in the foreground and color-coded as before.}
    \label{fig:local_ass_cw}
\end{figure*}

A useful measure describing the tendency of individuals in a network to link to others with similar properties or attributes is assortativity \cite{newman2003}, which yields one assortativity coefficient for a whole network. An assortativity coefficient of $r=1$ means that all edges in the network only connect nodes of the same type, while for $r=-1$, the edges only connect nodes of different type (hence, the network is strongly disassortative). It has been argued that such a global view might obstruct insights into local differences between individuals or groups \cite{peel2018}. Using local assortativity has been proposed in order make those differences visible---each node $l$ in a network is assigned a local assortativity score $r(l)$, such that differences in the score can be compared across all nodes. Local assortativity $r(l)$ is defined by the equation
\cite{peel2018} 
\begin{eqnarray}
r(l) = \frac{1}{Q_{max}} \sum_g (e_{gg}(l) - a_{g}b_{g}),
\end{eqnarray}

with $Q_{max}$ as the maximum modularity, which normalizes the assortativity coefficient. Maximum modularity is reached if all edges in the network only connect nodes of the same type.
The proportion of edges in the local neighborhood of node $l$ which connect nodes of the same type $g$ is compared to $a_{g}b_{g}$, which is the proportion of edges between nodes of group $g$ if one would randomly create edges between nodes, while keeping the total number of outgoing and incoming edges for each type constant. $a_{g}$ here is the proportion of edges starting from nodes of group $g$, while $b_{g}$ is the proportion of edges ending at nodes of group $g$.
In general, $e_{gh}$ is given by
\begin{eqnarray}
e_{gh}=\sum_{i:y_{i}=g} \sum_{j:y_{j}=h} w(i;l) \frac{A_{ij}}{k_{i}}.
\end{eqnarray}
$w(i;l)$ is a distribution over all nodes designed to capture the mixing patterns within the local neighborhood of node $l$. We follow \cite{peel2018} in choosing the personalized PageRank vector. The local assortativity coefficient is also capable of including incomplete metadata---since in the current data set, not all users could be classified, this feature is beneficial. In the histograms displayed in Fig~\ref{fig:local_ass_el} and Fig~\ref{fig:local_ass_cw}, the node contributions to the histograms were adjusted according to the weight 
\begin{eqnarray}
z_{l} = \sum_{gh} e_{gh}(l),
\end{eqnarray}
the sum of local edge counts with \textit{known} metadata. (Code available at \cite{peelgit}.)

In Fig~\ref{fig:local_ass_el} and Fig~\ref{fig:local_ass_cw}, the local assortativity distributions of users of the different groups are displayed. The distributions are multimodal, i.e. exhibit more than one peak. In the election data set, users of the majority cluster (A) have their largest peak at a local assortativity close to $1$, i.e. they reply mainly to users of their own clusters (and, since local assortativity also takes into account the assortativity of the broader neighborhood, also mainly interact with users who do the same). A second, yet smaller peak is visible with negative local assortativity (around $-0.3$). Hence, some majority cluster users mainly seek debate with users from other clusters.
The reverse holds for the minority cluster (B): There, most of the users reply to users from the other groups. A smaller peak is visible at positive local assortativity (around $0.6$). Users from the intermediate region (C) also exhibit a multimodal distribution, with one slightly negative peak and one peak at around $0.6$. 
The NYE data shows an even more pronounced trend: Only few users of the minority cluster get assigned a positive local assortativity. Majority group users again show a bimodal distribution with one peak close to $r_{l}=1$, the other peak is now at around $r_{l}=-0.6$.


Further insight into the interaction patterns between the different clusters is provided by Table \ref{tab:ia_patterns_el}, which shows counts of replies between and within the clusters. 
\begin{table*}[!ht]
\small

\centering
\caption{Reply interactions between and within the different clusters---the columns show what cluster the replies are from, the rows who the cluster replies to. The percentages (in brackets) are given with respect to the overall number of replies a cluster has given. Note that the column entries do not sum up to the overall number of given replies since there are also replies from users that do not show up in the retweet networks, hence could not be classified.} \label{tab:ia_patterns_el}
\vspace{0.3cm}
\begin{adjustwidth}{-1.25in}{0in}

\begin{tabular}{llll|lll}
\hline
& \multicolumn{3}{c}{Election} & \multicolumn{3}{c}{NYE}\\
to/from&  Majority cl. &  Minority cl. & Intermediate cl. & Majority cl. &  Minority cl. &Intermediate cl. \\ 
\hline
Majority cl.  &          18,185 (59.4\%)&   10,590 (44.5\%) & 3,596 (53.0\%)  &    3,234 (50.5\%) &    2,814 (57.9\%) & 821 (66.5\%) \\
Minority cl. &    5,778 (18.9\%) &  5,964 (25.1\%) &    1,035 (15.3\%)   & 1,445 (22.6\%) &  827 (17.0\%) &   
162 (13.1\%) \\
Intermediate cl. &    4,770 (15.6\%) &    6,025 (25.3\%)  &        1,757 (25.9\%)  &    1,041 (16.3\%)  &     949 (19.5\%) & 191 (15.5\%)   \\ \hline
Overall &      30,615 &   23,790 &  6,785 &    6,403 &    4,858  &       1,235 \\
\hline
\end{tabular}
\end{adjustwidth}
\end{table*}
While users of the majority cluster of the retweet network address almost 60\% (election) or 50\% (NYE) of their replies to other users of their own cluster, the opposite holds for the other groups: Most of their replies are directed towards users from the majority cluster. For the election, users of the intermediate cluster replied more often to others from their cluster than to the minority pole, while for the minority pole, replies to the own and replies to the intermediate cluster were almost equal in number. The number of tweets from intermediate users was considerably lower compared to tweets from the poles, hence it appears that users of the minority pole were motivated to challenge both intermediate and majority pole users. A similar tendency can be found for the NYE incident. 

To sum up, interaction patterns within and between the different groups are heterogeneous: Some users from every group seek debate with the differently-minded and others show a tendency to discuss amongst their own cluster. Nevertheless, the minority pole by far shows a stronger tendency to reply to users from different clusters than the majority pole.

\section*{Discussion}
Vincent Price's comparison of public debate and town meetings receives a refinement in the conclusion of \cite{price1992}. He states that: 
\begin{quote}
    ``The democratic foundations of the concept of public opinion are indisputable; far less so are the democratic foundations of day-to-day political decisions, even when they are formed out of public debate. [...] We may well compare public debate to a town meeting---provided we keep in mind that although some town meetings enjoy free-flowing debate, there are other meetings for which almost no one shows up, at which powerful leaders and organized coalitions dominate, and at which people with minority viewpoints are shouted down or left standing outside.'' \cite[~p. 91]{price1992}

\end{quote}

The paragraph is quoted at length here since it illustrates aspects of public debate that might be amplified in online environments, specifically by social media. While the facilitation of communication bears the potential of enabling minorities and their concerns to gain public attention \cite{toepfl2015,fraser1990}, it can easily introduce systematic biases in the perception of public opinion. 
Online user comments are generally an important source of information for many and help in judging whether a product is good, a certain video is worth watching---or whether certain views should be taken into consideration in political decisions. While the experiences and opinions of others can provide a very useful basis for decision in these contexts, naive reliance on what others express online can be collectively dangerous, especially in an era in which social media shapes politics to an unprecedented extent. If groups with certain minority opinions manage to become increasingly visible, their opinions might appear more socially acceptable and accepted than they actually are. Such disparities can be problematic since perceived public opinion, as different studies have shown, can have a persuasive and/or silencing effect \cite{noelle1984,noelle2004,lee2010,noelle1980,neubaum2017monitoring}. Under certain circumstances, committed minorities might even be able to gain public predominance, while the majority falls silent \cite{gaisbauer2019,banisch2020social}. In the context of Twitter, this phenomenon might be especially problematic due to its tight links with traditional media and news outlets, where Twitter content is often directly taken as representing public opinion \cite{mcgregor2019social} or used as a source in routine coverage  \cite{paulussen2014,broersma2013,mcgregor2020twitter}, introducing either strongly biased data or leading to a potentially unjustified alarmism in coverage.

This explorative study has two main findings:
\begin{itemize}
\item Disproportionately many replies come from users which constitute, if retweet networks are considered, a minority---composed of accounts by right to far-right parties, politicians, and users retweeting their content. It is hence probable that the content produced by this group influenced mere observers' perceptions of public opinion to a degree that did not reflect their real number.
\item Users of different clusters also diverge in who they tend to reply to. While users from the majority cluster tend to interact mainly amongst themselves, users from the intermediate and especially from the minority cluster are more keen on confronting differently-minded others (Fig~\ref{fig:local_ass_el}, Fig~\ref{fig:local_ass_cw}, Table \ref{tab:ia_patterns_el}). 
\end{itemize} 
It is hence shown that users retweeting mostly right-wing populist contents show a stronger willingness to express themselves in the form of replies in both case studies. First-order replies are nearly identical in number with those from users retweeting mostly left-leaning parties, politicians and contents. While users retweeting center to center-right parties show similar activity patterns as left-leaning ones, they are comparatively small in number, which might reflect that this voter group rarely uses Twitter.
Public opinion in Noelle-Neumann's sense, assessed through reply sections, appears balanced between the two poles, while retweet networks suggest that the actual opinion proportions on the platform are very different. Noelle-Neumann also states that future expectations---i.e., that one opinion camp might gain ground in the future---might already increase the willingness of opinion expression now \cite{noelle2004}. This is in line with the fact that right-wing populist movements have gained strength in recent years in Germany. 

Previous findings parallel the results of this work:
In a study from Switzerland, it has been shown that users with a right-wing political leaning engage more frequently in the comment sections of news pages \cite{friemel2015}, an effect also visible in Table \ref{tab:active perc} and in \cite{toepfl2015}, where comment sections are interpreted as counter-public spaces. As an explanation for this effect in connection to the rise of right-wing populism, Schweiger \cite{schweiger2017informierte} argues that misperceptions of the opinion climate, fuelled by news consumption through social media, lead to higher willingness of opinion expression for certain social groups---especially for those who already put less trust in established media sources, and those who lack awareness about the biases implicit in social media. The proposed method here could contribute to a substantiation of these claims and findings in subsequent studies.

With the proposed method, differences in willingness of opinion expression can be made visible. We stress that the method proposed here is not restricted to the specific cases studies. With a suitable seed set of users, any debate on Twitter can be analysed analogously. It is, nevertheless, limited in scope: It attempts to gain a comprehensive structural view on Twitter debate, but does not analyze the content of the tweets. Moreover, a proportion of users in the data sets remains unclassified by the employed method since they did not appear in the retweet networks. Complementary methods of classification should be sought.

Finally, Twitter is only one social media platform, which in addition is not representative of the general population \cite{mellon2017twitter}. A potentially insightful avenue for future inquiry might be the comparison of reply sections with comment sections of online newspapers \cite{willaert2019facilitating}. Still, as we have argued, both Twitter's platform design as well as its echo in traditional media outlets at least implicitly award Twitter the role of the host of the big town meeting called public debate. We therefore deem it increasingly important to develop methods which enable a better understanding of which viewpoints are prominently featured on the platform, and which ones remain mostly unspoken or unheard.

\section*{Acknowledgments}
This project has received funding from the European Union's Horizon 2020 research and innovation programme under Grant Agreement No. 732942 (www.ODYCCEUS.eu). We are grateful for the repeated discussion of the ideas described in the paper with Stephanie Juetersonke, Stephan Poppe, Wolfram Barfuss, Roger Berger, Marcel Sarkoezy and the members of \textsc{Odycceus}.

% Either type in your references using
% \begin{thebibliography}{}
% \bibitem{}
% Text
% \end{thebibliography}
%
% or
%
% Compile your BiBTeX database using our plos2015.bst
% style file and paste the contents of your .bbl file
% here. See http://journals.plos.org/plosone/s/latex for 
% step-by-step instructions.
% 


\begin{thebibliography}{10}

\bibitem{jungherr2015}
Jungherr A.
\newblock Analyzing political communication with digital trace data.
\newblock Cham, Switzerland: Springer. 2015;.

\bibitem{bruns2014crisis}
Bruns A, Burgess J.
\newblock Crisis communication in natural disasters: The Queensland floods and
  Christchurch earthquakes.
\newblock Twitter and society [Digital Formations, Volume 89]:. 2014; p.
  373--384.

\bibitem{mellon2017twitter}
Mellon J, Prosser C.
\newblock Twitter and Facebook are not representative of the general
  population: Political attitudes and demographics of British social media
  users.
\newblock Research \& Politics. 2017;4(3):2053168017720008.

\bibitem{mislove2011}
Mislove A, Lehmann S, Ahn YY, Onnela JP, Rosenquist JN.
\newblock Understanding the demographics of Twitter users.
\newblock In: Fifth international AAAI conference on weblogs and social media;
  2011.

\bibitem{broersma2013}
Broersma M, Graham T.
\newblock Twitter as a news source: How Dutch and British newspapers used
  tweets in their news coverage, 2007--2011.
\newblock Journalism practice. 2013;7(4):446--464.

\bibitem{paulussen2014}
Paulussen S, Harder RA.
\newblock Social media references in newspapers: Facebook, Twitter and YouTube
  as sources in newspaper journalism.
\newblock Journalism practice. 2014;8(5):542--551.

\bibitem{mcgregor2019social}
McGregor SC.
\newblock Social media as public opinion: How journalists use social media to
  represent public opinion.
\newblock Journalism. 2019;20(8):1070--1086.

\bibitem{mcgregor2020twitter}
McGregor SC, Molyneux L.
\newblock Twitter’s influence on news judgment: An experiment among
  journalists.
\newblock Journalism. 2020;21(5):597--613.

\bibitem{noelle1984}
Noelle-Neumann E. Public Opinion. Our Social Skin; 1984.

\bibitem{price1992}
Price V.
\newblock Public opinion. vol.~4; 1992.

\bibitem{bruns2015twitter}
Bruns A, Burgess J.
\newblock Twitter hashtags from ad hoc to calculated publics.
\newblock Hashtag publics: The power and politics of discursive networks. 2015;
  p. 13--28.

\bibitem{trilling2015}
Trilling D.
\newblock Two different debates? Investigating the relationship between a
  political debate on TV and simultaneous comments on Twitter.
\newblock Social science computer review. 2015;33(3):259--276.

\bibitem{gearhart2014}
Gearhart S, Kang S.
\newblock Social media in television news: The effects of Twitter and Facebook
  comments on journalism.
\newblock Electronic News. 2014;8(4):243--259.

\bibitem{schweiger2008}
Schweiger W, Weiherm{\"u}ller M.
\newblock {\"O}ffentliche Meinung als Online-Diskurs--ein neuer empirischer
  Zugang.
\newblock Publizistik. 2008;53(4):535--559.

\bibitem{tumasjan2010}
Tumasjan A, Sprenger TO, Sandner PG, Welpe IM.
\newblock Predicting elections with twitter: What 140 characters reveal about
  political sentiment.
\newblock In: Fourth international AAAI conference on weblogs and social media;
  2010.

\bibitem{burnap2016ukelection}
Burnap P, Gibson R, Sloan L, Southern R, Williams M.
\newblock 140 characters to victory?: Using Twitter to predict the UK 2015
  General Election.
\newblock Electoral Studies. 2016;41:230--233.

\bibitem{habermas2015}
Habermas J.
\newblock Between facts and norms: Contributions to a discourse theory of law
  and democracy.
\newblock John Wiley \& Sons; 2015.

\bibitem{noelle2004}
Noelle-Neumann E, Petersen T.
\newblock The spiral of silence and the social nature of man.
\newblock In: Handbook of political communication research. Routledge; 2004. p.
  357--374.

\bibitem{springer2018}
Springer N, K{\"u}mpel AS.
\newblock User-Generated (Dis) Content.
\newblock In: Journalismus im Internet. Springer; 2018. p. 241--271.

\bibitem{lee2010}
Lee EJ, Jang YJ.
\newblock What do others’ reactions to news on Internet portal sites tell us?
  Effects of presentation format and readers’ need for cognition on reality
  perception.
\newblock Communication research. 2010;37(6):825--846.

\bibitem{lee2012}
Lee EJ.
\newblock That's not the way it is: How user-generated comments on the news
  affect perceived media bias.
\newblock Journal of Computer-Mediated Communication. 2012;18(1):32--45.

\bibitem{friemel2015}
Friemel TN, D{\"o}tsch M. Online reader comments as indicator for perceived
  public opinion; 2015.

\bibitem{toepfl2015}
Toepfl F, Piwoni E.
\newblock Public spheres in interaction: Comment sections of news websites as
  counterpublic spaces.
\newblock Journal of Communication. 2015;65(3):465--488.

\bibitem{kaiser2017}
Kaiser J.
\newblock Public spheres of skepticism: Climate skeptics’ online comments in
  the German networked public sphere.
\newblock International Journal of Communication. 2017;11.

\bibitem{fraser1990}
Fraser N.
\newblock Rethinking the public sphere: A contribution to the critique of
  actually existing democracy.
\newblock Social text. 1990;(25/26):56--80.

\bibitem{baker_1990}
Baker KM.
\newblock In: Public opinion as political invention. Inventing the French
  Revolution: Essays on French Political Culture in the Eighteenth Century.
  Cambridge University Press; 1990. p. 167–200.

\bibitem{kalogeropoulos2017}
Kalogeropoulos A, Negredo S, Picone I, Nielsen RK.
\newblock Who shares and comments on news?: A cross-national comparative
  analysis of online and social media participation.
\newblock Social media+ society. 2017;3(4).

\bibitem{mustafaraj2011}
Mustafaraj E, Finn S, Whitlock C, Metaxas PT.
\newblock Vocal minority versus silent majority: Discovering the opionions of
  the long tail.
\newblock In: 2011 IEEE Third International Conference on Privacy, Security,
  Risk and Trust and 2011 IEEE Third International Conference on Social
  Computing. IEEE; 2011. p. 103--110.

\bibitem{ross1977false}
Ross L, Greene D, House P.
\newblock The “false consensus effect”: An egocentric bias in social
  perception and attribution processes.
\newblock Journal of experimental social psychology. 1977;13(3):279--301.

\bibitem{mullen1992group}
Mullen B, Dovidio JF, Johnson C, Copper C.
\newblock In-group-out-group differences in social projection.
\newblock Journal of Experimental Social Psychology. 1992;28(5):422--440.

\bibitem{mdr}
Reissing C. Sachsens CDU-Kandidaten schließen Koalition mit AfD aus; 2019.
\newblock
  https://www.mdr.de/nachrichten/politik/regional/cdu-schliesst-koalition-afd-aus-sachsen-100.html
  (last accessed: 20 June 2020).

\bibitem{tagesspiegel}
Fiedler M, Jansen F. Was geschah an Silvester in Leipzig-Connewitz?; 2020.
\newblock
  https://www.tagesspiegel.de/politik/angriff-auf-polizisten-wirft-fragen-auf-was-geschah-an-silvester-in-leipzig-connewitz/25386832.html
  (last accessed: 20 June 2020).

\bibitem{Conover2011}
Conover MD, Goncalves B, Ratkiewicz J, Flammini A, Menczer F.
\newblock Predicting the Political Alignment of Twitter Users.
\newblock In: 2011 IEEE Third International Conference on Privacy, Security,
  Risk and Trust and 2011 IEEE Third International Conference on Social
  Computing; 2011. p. 192--199.

\bibitem{conover2011polarization}
Conover MD, Ratkiewicz J, Francisco M, Gon{\c{c}}alves B, Menczer F, Flammini
  A.
\newblock Political polarization on twitter.
\newblock In: Fifth international AAAI conference on weblogs and social media;
  2011.

\bibitem{Gaumont2018}
Gaumont N, Panahi M, Chavalarias D.
\newblock Reconstruction of the socio-semantic dynamics of political activist
  Twitter networks—Method and application to the 2017 French presidential
  election.
\newblock PLOS ONE. 2018;13(9):1--38.
\newblock doi:{10.1371/journal.pone.0201879}.

\bibitem{jacomy2014}
Jacomy M, Venturini T, Heymann S, Bastian M.
\newblock ForceAtlas2, a continuous graph layout algorithm for handy network
  visualization designed for the Gephi software.
\newblock PloS one. 2014;9(6).

\bibitem{bastian2009gephi}
Bastian M, Heymann S, Jacomy M.
\newblock Gephi: an open source software for exploring and manipulating
  networks.
\newblock In: Proceedings of the International AAAI Conference on Web and
  Social Media. vol.~3; 2009.

\bibitem{noack2009}
Noack A.
\newblock Modularity clustering is force-directed layout.
\newblock Physical Review E. 2009;79(2):026102.

\bibitem{venturini2019}
Venturini T, Jacomy M, Jensen P. What do we see when we look at networks; 2019.

\bibitem{sousa2010}
Sousa D, Sarmento L, Mendes~Rodrigues E.
\newblock Characterization of the twitter@ replies network: are user ties
  social or topical?
\newblock In: Proceedings of the 2nd international workshop on Search and
  mining user-generated contents; 2010. p. 63--70.

\bibitem{aragon2013}
Arag{\'o}n P, Kappler KE, Kaltenbrunner A, Laniado D, Volkovich Y.
\newblock Communication dynamics in twitter during political campaigns: The
  case of the 2011 Spanish national election.
\newblock Policy \& internet. 2013;5(2):183--206.

\bibitem{yardi2010}
Yardi S, boyd d.
\newblock Dynamic debates: An analysis of group polarization over time on
  twitter.
\newblock Bulletin of science, technology \& society. 2010;30(5):316--327.

\bibitem{nuernbergk2016}
Nuernbergk C, Conrad J.
\newblock Conversations and campaign dynamics in a hybrid media environment:
  Use of Twitter by members of the German Bundestag.
\newblock Social Media+ Society. 2016;2(1).

\bibitem{newman2003}
Newman ME.
\newblock Mixing patterns in networks.
\newblock Physical Review E. 2003;67(2).

\bibitem{peel2018}
Peel L, Delvenne JC, Lambiotte R.
\newblock Multiscale mixing patterns in networks.
\newblock Proceedings of the National Academy of Sciences.
  2018;115(16):4057--4062.

\bibitem{peelgit}
Peel L. MultiscaleMixing; 2018.
\newblock Available from: \url{github.com/piratepeel/multiscalemixing}.

\bibitem{noelle1980}
Noelle-Neumann E.
\newblock Die Schweigespirale. Öffentliche Meinung--Unsere soziale Haut.
\newblock Riper [ie Piper]; 1980.

\bibitem{neubaum2017monitoring}
Neubaum G, Kr{\"a}mer NC.
\newblock Monitoring the opinion of the crowd: Psychological mechanisms
  underlying public opinion perceptions on social media.
\newblock Media Psychology. 2017;20(3):502--531.
\newblock doi:{10.1080/15213269.2016.1211539}.

\bibitem{gaisbauer2019}
Gaisbauer F, Olbrich E, Banisch S.
\newblock Dynamics of opinion expression.
\newblock Physical Review E. 2020;102(4):042303.

\bibitem{banisch2020social}
Banisch S, Gaisbauer F, Olbrich E.
\newblock How social feedback processing in the brain shapes collective opinion
  processes in the era of social media.
\newblock arXiv:200308154. 2020;.

\bibitem{schweiger2017informierte}
Schweiger W.
\newblock Der (des)informierte B{\"u}rger im Netz.
\newblock Springer; 2017.

\bibitem{willaert2019facilitating}
Willaert T, Banisch S, Van~Eecke P, Beuls K.
\newblock Facilitating on-line opinion dynamics by mining expressions of
  causation. The case of climate change debates on The Guardian.
\newblock arXiv:191201252. 2019;.

\end{thebibliography}
\end{document}